\documentclass[useAMS,usenatbib]{mn2e}
\usepackage{amssymb}
\usepackage{amsmath}
\usepackage{times}
\usepackage{color}
\usepackage{graphicx}
\usepackage[T1]{fontenc}
\usepackage{aecompl}

\title[A statistical analysis of BGGs and BCGs in the last 3 billion years]
  {Galaxy And Mass Assembly (GAMA): Testing galaxy formation models through the most massive galaxies in the Universe}
\author[P. Oliva-Altamirano et al.]
  {P.~Oliva-Altamirano$^{1,2}$\thanks{E-mail:poliva@astro.swin.edu.au},
  S.~Brough$^2$, C.~Lidman$^2$, W.~J.~Couch$^{1,2}$, A.~M.~Hopkins$^2$, \newauthor M.~Colless$^{3}$, E.~Taylor$^4$, A.~S.~G.~Robotham$^{5,6}$, M.~L.~P. Gunawardhana$^{2,7,8}$, \newauthor T.~Ponman$^9$,  I.~Baldry$^{10}$, A.~E.~Bauer$^2$, J.~Bland-Hawthorn $^7$, M. Cluver$^{2,11}$, \newauthor E.~Cameron$^{12}$,  C.~J.~Conselice$^{13}$, S. Driver$^{5,6}$, A.~C.~Edge$^{14}$, A.~W.~Graham$^{1}$, \newauthor E.~van Kampen$^{15}$, M.~A.~Lara-L\' opez$^2$, J.~Liske$^{15}$, A.~R.~L\' opez-S\' anchez$^2$,  J.~Loveday$^{16}$, \newauthor  S.~Mahajan$^{17}$, J. Peacock$^{18}$,  S.~Phillipps$^{19}$, K.~A.~Pimbblet$^{20,21,22}$, R.~G.~Sharp$^{3}$  
\\% starts a new line in the
   $^1$Centre for Astrophysics \& Supercomputing, Swinburne University of Technology, Hawthorn 3122, Australia\\$^2$Australian Astronomical Observatory, PO Box 915, North Ryde, NSW 1670, Australia\\$^3$Research School of Astronomy \& Astrophysics, The Australian National University, Mount Stromlo Observatory, ACT 2611, Australia\\$^4$School of Physics, The University of Melbourne, Parkville 3010, Australia\\$^5$International Centre for Radio Astronomy Research (ICRAR), University of Western Australia, Crawley, WA 6009, Australia\\$^6$School of Physics \& Astronomy, University of St Andrews, North Haugh, St Andrews KY16 9SS, UK.\\$^7$School of Physics University of Sydney, Sydney Institute for Astronomy, NSW 2006 Australia\\$^8$Institute for Computational Cosmology, Department of Physics, Durham University, South Road, Durham DH1 3 LE, U.K\\$^9$School of Physics and Astronomy, University of Birmingham, Edgbaston, Birmingham B15 2TT, UK\\$^{10}$Astrophysics Research Institute, Liverpool John Moores University, IC2, Liverpool Science Park, 146 Brownlow Hill, Liverpool, L3 5RF \\$^{11}$Department of Astronomy, University of Cape Town, Private Bag X3, Rondebosch, 7701, South Africa\\$^{12}$Max Planck Institute for Nuclear Physics (MPIK), Saupfercheckweg 1, D-69117 Heidelberg, Germany\\$^{13}$School of Physics \& Astronomy, University of Nottingham, University Park, Nottingham NG7 2RD, UK\\$^{14}$Institute of Computational Cosmology, University of Durham, Durham, DH1 3LE, UK\\$^{15}$European Southern Observatory, Karl-Schwarzschild-Str. 2, D-85748 Garching, Germany\\$^{16}$Astronomy Centre, University of Sussex, Falmer, Brighton BN19QH, UK\\$^{17}$School of Mathematics and Physics, University of Queensland, Brisbane, QLD 4072 Australia\\$^{18}$Institute for Astronomy Royal Observatory, Blackford Hill, Edinburgh EH9 3HJ\\$^{19}$Department of Physics, University of Bristol, Tyndall Avenue, Bristol BS8 1TL, UK\\$^{20}$School of Physics, Monash University, Clayton, VIC 3800, Australia\\$^{21}$Department of Physics, University of Oxford, Denys Wilkinson Building, Keble Road, Oxford OX1 3RH, UK\\$^{22}$Department of Physics and Mathematics, University of Hull, Cottingham Road, Hull, HU6 7RX, UK}
   
\date{Released 2002 Xxxxx XX}
\pagerange{\pageref{firstpage}--\pageref{lastpage}} \pubyear{2002}
\def\LaTeX{L\kern-.36em\raise.3ex\hbox{a}\kern-.15em
    T\kern-.1667em\lower.7ex\hbox{E}\kern-.125emX}

\begin{document}
\label{firstpage}
\maketitle

\begin{abstract}

We have analysed the growth of Brightest Group Galaxies and Brightest Cluster Galaxies (BGGs/BCGs) over the last 3 billion years using a large sample of 883 galaxies from the Galaxy And Mass Assembly Survey. By comparing the stellar mass of BGGs and BCGs in groups and clusters of similar dynamical masses, we find no significant growth between redshift $z=0.27$ and $z=0.09$. We also examine the number of BGGs/BCGs that have line emission, finding that approximately 65 per cent of BGGs/BCGs show H$\alpha$ in emission. From the galaxies where the necessary spectroscopic lines were accurately recovered (54 per cent of the sample), we find that half of this (i.e. 27 per cent of the sample) harbour on-going star formation with rates up to $10\,$M$_{\odot}$yr$^{-1}$, and the other half (i.e. 27 per cent of the sample) have an active nucleus (AGN) at the centre. BGGs are more likely to have ongoing star formation, while BCGs show a higher fraction of AGN activity.  By examining the position of the BGGs/BCGs with respect to their host dark matter halo we find that around 13 per cent of them do not lie at the centre of the dark matter halo. This could be an indicator of recent cluster-cluster mergers. We conclude that BGGs and BCGs acquired their stellar mass rapidly at higher redshifts as predicted by semi-analytic models, mildly slowing down at low redshifts.  
\end{abstract}

\begin{keywords}
 galaxies: clusters: general -- galaxies: elliptical and lenticular, cD -- galaxies: evolution -- galaxies: groups: general -- galaxies: haloes -- galaxies: star formation. 
\end{keywords}

%============================================================================================
\section{Introduction}
In the hierarchical model of structure formation, galaxies grow in size and stellar mass by accreting other galaxies and material from their surroundings. The brightest group and cluster galaxies (hereafter BGGs and BCGs) are the most extreme examples of this process and the most luminous objects known at the present epoch. Their properties are shown to be different to other early-type galaxies in clusters and groups \citep[e.g. higher velocity dispersions than any other elliptical of the same mass, extended light profiles, and systematically brighter than what would be inferred from the luminosity function;][]{LOH06,LINDEN07,SHEN13}. The uniqueness of their properties has been attributed to their privileged location at the centre of their host group or cluster \citep{HAUSMAN78}. However, their properties also correlate with the mass of their host cluster \citep[e.g.][]{COLLINS98,BURKE00,BROUGH02}.
\\\\
Most contemporary models of galaxy formation are based on the hierarchical assembly of dark matter halos \citep[][]{TOOMRE77,WHITE78} in the $\Lambda$CDM cosmology. In this paradigm, galaxies form at the centre of the halos. While N-body models accurately describe how the dark matter halos evolve, we are unable to simulate in detail the processes that lead to galaxy evolution. Halo abundance-matching (dark-matter-only) simulations and semi-analytical models (SAM) are two approaches that are commonly used in the literature. Halo abundance-matching models follow the behaviour of the $\Lambda$CDM cosmology with gravity \citep[e.g.][]{MOSTER13, LAPORTE13}. SAMs \citep[e.g.][]{DELUCIA07, TONINI12} are a combination of N-body simulations and analytic descriptions of galaxy formation physics \citep[i.e. star formation, dust extinction, AGN feedback, etc; for a review see][]{MUTCH13}. 
BCGs are particularly difficult to reproduce using these models, with their photometric colours tending to be bluer compared to observations, and their masses over-estimated \citep{BOWER06, CROTON06, DELUCIA07,COLLINS09,SILK12}.
\\\\ 
There are few models that focus solely on BGGs and BCGs. Among them we find \citet{DELUCIA07} who used the Millennium N-body simulation of \citet{SPRINGEL05} to model the development of BCGs over cosmic time. A similar  approach was used by \citet{TONINI12}.  More recently, \citet{MOSTER13} introduced an abundance-matching simulation of galaxy groups and clusters, using a statistical model constrained by observations. However, these simulations have not yet converged with observations. Despite these efforts, the assembly history and evolution of BGGs and BCGs is still poorly understood. SAMs and abundance-matching simulations predict a factor of three increase in the BCG stellar mass since $z=1$. On the other hand, earlier observational studies implied a very different result, arguing that BCG stellar masses 9 billion years ago were not very different to their stellar masses now \citep{ARAGON98,BAUGH99,STOTT08,STOTT10,COLLINS09}. 
\\\\
Recently, \citet{LIDMAN12} added new photometry from near-infrared imaging of clusters at $0.8<z<1.6$ to previous observations \citep[][]{STOTT08,STOTT10}. Their results show that from $0.1<z<0.9$, BCGs grow in mass by a factor of $1.8\pm0.3$. This is in closer agreement with the predictions from SAMs \citep{DELUCIA07,TONINI12}. Also \citet{LIN13}, with a sample from the \textit{Spitzer} IRAC Shallow Cluster Survey  \citep[ISCS;][]{EISENHARDT08} in the cluster mass range, $(2.4-4.5)\times 10^{14}\,$M$_{\odot}$, found remarkably good agreement  with the SAM of \citet{GUO11} over the redshift interval $0.5<z<1.5$ (growth by a factor of 2.3). However, below $z=0.5$, they found that the growth stalls, a result that is not seen in the models. In this paper we will examine this low redshift interval more closely ($0.09\leq z \leq 0.27$). 
\\\\
Another important prediction made by hierarchical formation models is that BCGs are assembling their mass through similar mass mergers with little gas present \citep{DELUCIA07,LAPORTE13}. However, recent analyses have shown that some BCGs harbour on-going star formation \citep{EDWARDS07,ODEA08,ODEA10,PIPINO09,LIU12,THOM12}. \citet{LIU12}, with an optical sample selected from the Sloan Digital Sky Survey \citep[SDSS][]{YORK00}, found that the star formation rate in BCGs is not always low, although it is not high enough to increase the stellar mass of the BCG by more than 1 per cent.  The existence of star formation in such old galaxies has important implications for simulations \citep{TONINI12}. 
\\\\
It is also generally assumed that the central galaxy of a cluster is also the brightest and most massive one (BCG). However, \citet[][]{BEERS83}, \citet[][]{POSTMAN95}, \citet[][]{LIN04}, \citet[][]{PIMBBLET06}, \citet{LINDEN07}, \citet[][]{BILDFELL08}, \citet[][]{HWANG08}, \citet[][]{SANDERSON09}, \citet[][]{COZIOL09}, and \citet[][]{SKIBBA11}  have demonstrated that this is not always the case. The proposed explanation for this is ongoing halo merging. Observationally, this can be understood as different stages in the hierarchical clustering process \citep{PIMBBLET08,BROUGH08}. 
\\\\
In this paper we analyse a sample of BGGs/BCGs selected from the Galaxy And Mass Assembly survey \citep[GAMA; ][]{DRIVER11}. Our sample is one of the largest accessible to date, covering a wide range of halo mass ($10^{12}-10^{15}\,$M$_{\odot}$). Throughout this paper we separate groups from clusters by a halo mass cut of M$_{\rm halo}=10^{14}\,$M$_{\odot}$\footnote{We tested our analysis by separating groups from clusters through multiplicity, and we find the results to be consistent.}. We analyse the BGG and BCG stellar mass growth spanning $0.27\leq z \leq 0.09$, and compare our results with the galaxy formation and evolution models of \citet{DELUCIA07}, \citet{TONINI12}, and \citet{MOSTER13}. In addition, we analyse the active galactic nucleus (AGN) and star formation (SF) activity within these galaxies as well as their position in their host halo. We are interested in: (a) exploring the impact of ongoing star formation on the growth of these giant galaxies, (b) comparing the properties of BGGs/BCGs that are not at the centre of the potential well with those that are, and (c) looking for correlations between the properties of BGGs/BCGs and the properties of the dark matter halos in which they live. Together, these will give us detailed information of the evolution history of group/clusters, and their brightest galaxies. 
\\\\ 
In Section 2 we describe the GAMA \citep{DRIVER11} survey and the different catalogues used in our analysis. In Section 3 we revisit the M$_*-$M$_{\rm halo}$ relationship. In Section 4 we describe our method for estimating the growth in BGGs/BCGs in the last 3 billion years. In Section 5 we analyse the BGG/BCG's AGN and Star formation activity. Section 6 examines the position of the BGGs/BCGs within their host halo. In Section 7 we discuss the main results obtained from our study, comparing them with model predictions. We present a summary of our final conclusions in Section 8. The cosmology adopted throughout this paper is H$_0$=70 kms$^{-1}$ Mpc$^{-1}$, $\Omega_M$=0.3, $\Omega_{\Lambda}$=0.7.
%============================================================================================

\section{The Galaxy And Mass Assembly (GAMA) survey}
The Galaxy And Mass Assembly (GAMA) survey is a multiwavelength galaxy survey \citep[][]{DRIVER11}. Beginning in 2008 it has obtained optical spectra from the 3.9m Anglo-Australian Telescope (AAT) in five regions of the sky covering $290\,$deg$^2$. GAMA~I contains three of these regions ($\sim144\,$deg$^2$). We specifically selected galaxies from this first phase of the survey \citep[][]{DRIVER11}.  There are $\sim 170,000$ galaxies in the GAMA~I sample, down to $r\sim 19.4$ mag in two regions each of $48\,$deg$^2$ and $r\sim19.8$ mag in a third region also of $48\,$deg$^2$. GAMA has a very high spectroscopic completeness \citep[on average 97 per cent;][]{DRIVER11}. This has been achieved by returning to each target area an average of 10 times \citep{ROBOTHAM10}. Sloan Digital Sky Survey \citep[SDSS;][]{YORK00} imaging has also been re-analysed for GAMA targets \citep{HILL11,KELVIN12}, making possible the stellar mass determinations from spectral energy distributions \citep[][]{TAYLOR11}. The spectra enables measurement of emission line star formation rates (SFR) form optical spectra \citep{GUNAWARDHANA13}. We work with those targets having reliable redshifts, i.e. quality value $nQ\geq3$ \citep{BALDRY10}.
\\\\
The group-finding, stellar mass and star formation rate measurements are crucial for our analysis. These are described in the following sections.
 %------------------------------------------------------------------------------------------------------------------------------------------------------------------
%***********************************************************************************
\begin{figure}
\includegraphics[width=\linewidth]{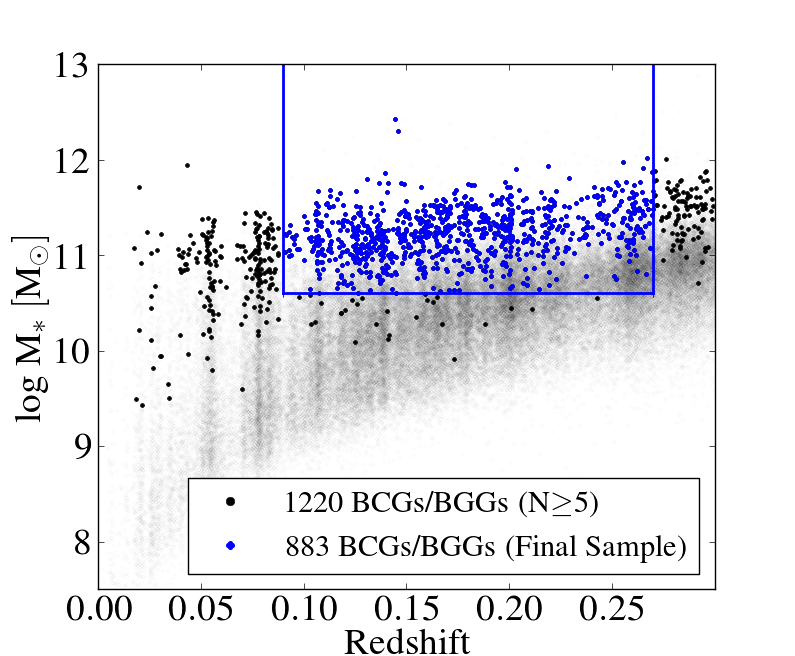}
\caption{Stellar mass as function of redshift, z. In grey dots we show the stellar mass of GAMA~I galaxies as a function of redshift. The M$_{*}$ threshold is a function of z. Galaxies in black are the BGGs/BCGs from halos with multiplicity greater than or equal to 5. The BGGs/BCGs in blue delimited by the blue box, are the final sample of 883 BCGs, they are taken from halos with multiplicity N~$\geq 5$ at redshifts $0.09<z<0.27$ and with M$_* \geq 3 \times 10^{10}\,$ M$_{\odot}$. }  
\label{fig:sample}
\end{figure}
%***********************************************************************************

\subsection{The GAMA Galaxy Group Catalogue (G$^3$C)}\label{sec:group}
A full description of the GAMA Galaxy Group Catalogue (G$^3$C) can be found in \citet{ROBOTHAM11}. We use version v04 of the group catalogue. The groups in this catalogue were identified with an adaptive friends-of-friends algorithm, tested with mock GAMA lightcones. These GAMA lightcones, have been constructed from $\Lambda$CDM N-body simulations \citep{SPRINGEL05} using the galaxy formation recipe of \citet{BOWER06}. In order to simulate realistic GAMA galaxies, the mock catalogues include the limitations of GAMA spectroscopy.  G$^3$C comprises 14,388 galaxy groups with multiplicity~$\geq 2$ containing 44,186 galaxies. This represents 40 per cent of the galaxies in GAMA. To prevent confusion, henceforth, all groups and clusters will be referred to as `halos' in the cases where distinction is not necessary.
\\\\
The methods to measure the halo properties in the G$^3$C were selected in order to be robust and  unbiased to perturbations even in groups with a small number of members. This was achieved by comparison with the mock catalogues. 
\\\\
It is crucial for this analysis to identify the BGG/BCG and the halo centre. In the group catalogue, the BCG is simply defined as the brightest (most luminous) galaxy in the halo in r$_{\rm AB}$ luminosity. To find the most suitable halo centre three definitions were tested: the centre of the total light of the halo (CoL), the iterative centre of light, or the brightest halo galaxy. The iterative centre of light was taken as the most robust definition due to the better match with mock halo positions.  The iterative centre of light produced a perfectly recovered centre position (i.e. within $3.5$ kpc) between observations and mocks 90 per cent of the time. This is significantly higher than $\sim70$ per cent of matches for the BGG/BCG centre, and $20$ per cent for the CoL method respectively. The iterative centre was also shown to be less sensitive to perturbations by individual members, and very stable as a function of multiplicity. For multiplicities $5<\,$N$\,<19$, the observed systems recover the position in the mock catalogues in ~88 per cent of the cases. The stability increases slightly for N$\,> 19$ (to 93 per cent). However, this should not affect this work as only 5 per cent of our systems contain more than 20 members ($\sim$50 clusters out of 883 systems). The procedure of finding the iterative centre of light consists of a number of iterations made in the halo r$_{\rm AB}$ luminosity. For each iteration the centroid of the halo r$_{\rm AB}$ luminosity is found, rejecting the distant galaxies and selecting the brightest from the remaining ones. The central galaxy was defined as that closest to the centroid of the overall light distribution of the system.
\\\\
 The halo velocity dispersions ($\sigma_{\rm halo}$) are measured as described in \citet{BEERS90}, and R$_{\rm 50}$ (the radius of the 50th percentile group member) was selected as the definition of the radius. From an accurate recovery of these two properties, the halo mass can be calculated following the virial theorem. The dynamical mass (M$_{\rm halo}$) is proportional to A$\sigma^2_{\rm halo}R_{\rm 50}$; where A is the scaling factor that leads to a median unbiased mass estimate. It is defined as A$\sim$M$_{\rm DM}$/M$_{\rm FoF}$, where M$_{\rm DM}$ is the mass of the dark matter halo extracted from the Millennium simulation (not dependent on the details of the semi-analytic models), and M$_{\rm FoF}$ is the mass of the group with members identified through friends-of-friends algorithm. The uncertainties in this method were tested through density distributions (log~$\sigma^{2}_{\rm FoF}/\sigma^{2}_{\rm DM} -\, $log~Rad$_{\rm 50-FoF}$/Rad$_{\rm 50-DM}$). The dynamical mass was shown to be more consistent in halos with larger multiplicities (N~$\geq 5$) since the radius is recovered more accurately in those cases. We therefore limit our analysis to BGGs/BCGs in halos with multiplicity N~$\geq 5$, resulting in 1220 halos (i.e. 1220 BGGs/BCGs).  The scaling factor in these cases  were shown to be $A=10$ from a global optimization of the match between the friends-of-friends groups and the mock catalogues. The G$^3$C covers a wide halo mass range $10^{12}\,$M$_{\odot}<\,$M$_{\rm halo}<10^{15}\,$M$_{\odot}$, which is used in this paper.  
\\\\ 
We also make use of other parameters available in the group catalogue including modality, dominance, and relative overdensity. The modality describes the Gaussianity of the velocity dispersion distribution in the halo, and is defined as (1+skewness$^2$)/(3+kurtosis$^2$). For Gaussian systems it is close to 1/3. The dominance is defined as the luminosity gap between the BGG/BCG and the second brightest galaxy in a halo ($\Delta$m$_{\rm 1,2}$). In G$^3$C the magnitudes used are apparent r$_{\rm AB}-\,$band magnitudes from SDSS. Finally the relative halo overdensity is a measure of how isolated the group is relative to larger scale structures. It is calculated by detecting the number of objects within a comoving cylinder of radius 1.5 h$^{-1}$Mpc and radial depth of 36 h$^{-1}$Mpc centred at the centre of the halo.
%------------------------------------------------------------------------------------------------------------------------------------------------------------------
%***********************************************************************************
\begin{figure}
\includegraphics[width=\linewidth]{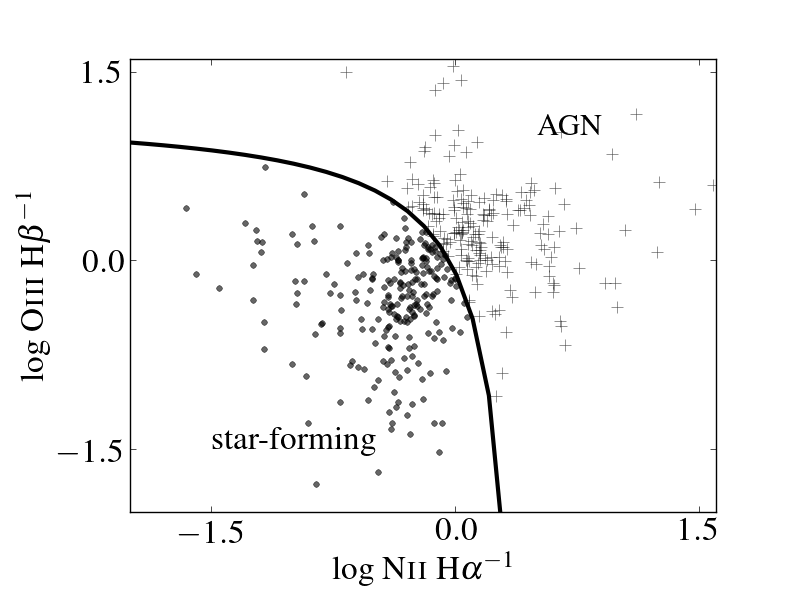}
\caption{BPT digram. The BGGs and BCGs are shown as grey crosses if they present AGN activity, and as black circles if they are star-forming.The solid line indicates the distinction between AGN and star-forming galaxies defined by \citep{KEWLEY01}.}
\label{fig:BPT}
\end{figure}
%***********************************************************************************

\subsection{GAMA Stellar Mass Catalogue}
The stellar masses (M$_*$) for all GAMA galaxies are calculated as described in \citet{TAYLOR11}. We use catalogue version v08. The stellar mass estimates were derived by fitting the Spectral Energy Distributions \citep[SEDs;][BC03]{BRUZUAL03} to SDSS \citep{YORK00} \textit{ugriz} imaging reprocessed by the GAMA team \citep{HILL11}. The dust obscuration law applied was that of \citet{CALZETTI00}, and a \citet{CHABRIER03} initial mass function (IMF) was assumed. To account for aperture effects, a correction based on the S\' ersic fit to the surface brightness profiles is applied to the stellar masses \citep[see][]{TAYLOR11,KELVIN12}. 
\\\\
GAMA~I galaxies have stellar masses ranging from $3\times10^7$ to $3\times10^{12}\,$M$_{\odot}$; this is illustrated in Figure~\ref{fig:sample} by the grey dots. The stellar masses of our sample of 1220 BCGs are between $1\times10^9$ and $3\times10^{12}\,$M$_{\odot}$, also shown in Figure~\ref{fig:sample} as black circles, these BCGs live in halos with N~$\, \geq5$. To ensure that we are not biased by mass-dependant selection effects in our analysis, we select a volume limited sample with a stellar mass limit $3\times10^{10}\,$M$_{\odot}$ over $0.09 \leq z \leq 0.27$ (blue circles within the box in Figure~\ref{fig:sample}). The stellar mass limit ensures the possibility of finding the `progenitors' of the lower redshift halos at higher redfshifts.
This final sample contains 883 halos (i.e. 883 BGGs/BCGs).  %------------------------------------------------------------------------------------------------------------------------------------------------------------------

\subsection{GAMA Emission Line catalogue}\label{sec:bpt}
GAMA spectra were obtained using the AAOmega multi-object spectrograph \citep{SHARP06}. The targets were observed with the 580V and 385R AAOmega gratings giving an observed wavelength range of $\sim$3700-8900~\AA~with a spectral resolution of 3.2~\AA~FWHM. The spectra are extracted, flat-fielded and wavelength and flux-calibrated as described in \citet{HOPKINS13}. Spectral line measurements are made assuming a single Gaussian approximation fitted from a common redshift value and line-width within adjacent sets of lines. The galaxies are classified as Active Galactic Nuclei (AGN) and Star Forming (SF) galaxies using the division described by \citet{KEWLEY01} in the \citet*[BPT;][]{BPT} diagram. This division is based on the [N$_{\rm II}$]/H$\alpha$ and [O$_{\rm III}$]/H$\beta$ line ratios, shown in Figure \ref{fig:BPT}. If measurements for all four lines are not available then the two-line diagnostic given by  \citet{KEWLEY01}  was used, and if measurements for two lines are not available then the galaxy was classified as uncertain \citep{GUNAWARDHANA13}.
\\\\
The calculation of the star formation rates (SFR) is described in \citet{GUNAWARDHANA13} and the catalogue version we use here is v04.10. SFRs are calculated from the H$\alpha$ luminosities using the relationship defined by \citet{KENNICUTT98}. This uses a \citet{SALPETER55} initial mass function (IMF)  and we translate it to  \citep{CHABRIER03} IMF to be consistent with the stellar mass catalogue using the relationship given by \citet{BALDRY03}. The GAMA H$\alpha$ flux limit is $2.5\times10^{-16}$ergs$^{-1}$cm$^{-2}$ which corresponds to a SFR of $0.1\,$M$_{\odot}$yr$^{-1}$ at $z=0.27$. This is the threshold to recognize H$\alpha$ as an emission line. Therefore in our sample we take star-forming galaxies to be those with $0.1<$~SFR~$ \leq 100\,$M$_{\odot}$yr$^{-1}$. Any measurement higher than $100\,$M$_{\odot}$yr$^{-1}$ is potentially unreliable. Galaxies affected by sky lines lying at the same wavelength as the emission lines were excluded from the final sample.  
%============================================================================================

%***********************************************************************************
\begin{figure}
\includegraphics[width=\linewidth]{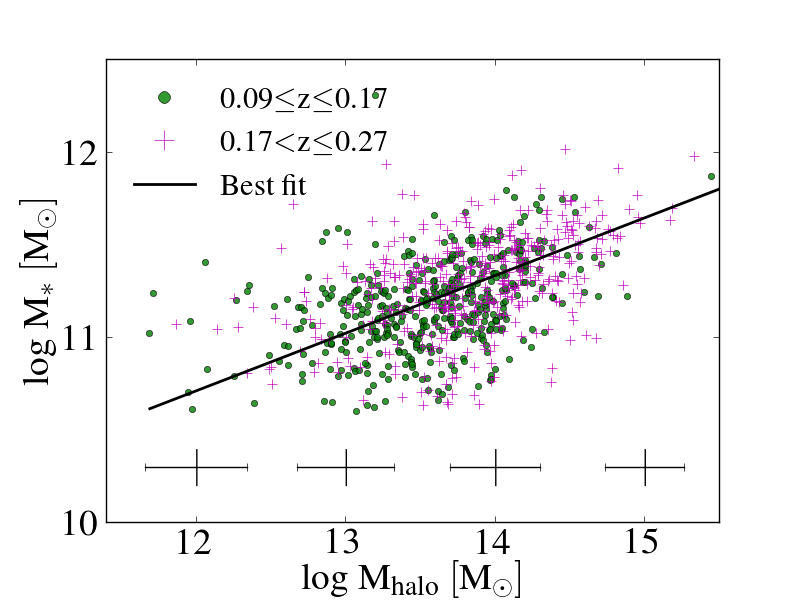}
\caption{BCG Stellar Mass~-~Host Halo Mass relationship. BGGs/BCGs at low-redshift (0.09~$\leq$~z~$\leq$~0.17) are shown as green circles and those at high-redshift (0.17~$<$~z~$\leq$~0.27) are shown as purple crosses. The best-fit relationship determined through a Bayesian approach (M$_* \propto$ M$_{\rm halo}^{0.32 \pm 0.09}$) is shown by the solid black line. The median M$_{\rm halo}$ and BCG M$_*$ error bars are plotted at the bottom of the figure. BGG/BCG stellar mass and host halo mass are correlated such that the halo grows faster than the BGG/BCG.}
\label{fig:fit}
\end{figure}
%***********************************************************************************

\section{BCG Stellar Mass$-$Halo Mass Relationship}
There is a known correlation between the stellar mass of the BCG and the mass of its host dark matter halo. The more massive the halo, the more massive the BCG. Many studies \citep[e.g.][]{ARAGON98, BROUGH05, BROUGH08, STOTT08, STOTT10,STOTT12, COLLINS09, HANSEN09, LIDMAN12} have explored the slope of this  M$_*-\,$M$_{\rm halo}$ correlation and have found it to be less than unity. This implies that the galaxy does not grow at the same rate as the cluster; the cluster acquires its mass faster than the BCG. 
\\\\ 
To study the M$_*-\,$M$_{\rm halo}$ relationship for the galaxies in our sample, we look for the best fit using Bayesian statistics. We treat the data as a 2D Gaussian with uncertainties in both variables, taking into account the intrinsic scatter. We generate a uniform prior to later maximise it (likelihood) through Markov Chain Monte-Carlo (MCMC) iteration. From the maximum posterior distribution the index of the power-law (hereafter referred as $b$ and M$_* \propto$ M$_{\rm halo}^{b}$) is found to be $b=0.32 \pm 0.09$. Our result is robust to flipping the axes. The goodness of the fit is tested through the efficiency of our MCMC implementation \citep{EMCEE12}. In Figure \ref{fig:fit} we show the best fit for the M$_*-\,$M$_{\rm halo}$ relationship as a solid line. We have divided the 883 BCGs into two redshift bins that are represented in this Figure by different colours. The low-redshift sample ($0.09 \leq z \leq 0.17$) is shown as green points, and the high-redshift sample ($0.17<z \leq 0.27$) as purple crosses. The blue error bars represent the median errors for each of the M$_*$ and M$_{\rm halo}$ bins. The power-law index of $\sim0.32$ implies that if the halo grows by a factor of 10 in dynamical mass, its BGG or BCG only gains a factor of two in stellar mass. 
\\\\
As a further analysis, we investigate the M$_*-\,$M$_{\rm halo}$ relationship for the different subsamples: (i) low vs high redshift, and (ii) groups vs clusters separated at M$_{\rm halo}=10^{14}\,$M$_{\odot}$. The power-law indexes for the low-redshift sample ($b=0.33 \pm 0.21$), and high-redshift sample ($b=0.30 \pm 0.20$) are consistent within the error bars, while the differences between groups and clusters is more apparent. The relationship is shallower for the groups than for the clusters, $b=0.19 \pm 0.20$, and $b=0.39 \pm 0.16$ respectively. However, the relationships of the groups and clusters are still consistent within the error bars. We do not find any significant change in the BGG/BCG stellar mass - host halo mass relationship as a function of redshift or halo mass.
\\\\
Comparing the index of the power-law to previous analyses we find that we are in good agreement with the analyses at similar redshift range, e.g. \citet{HANSEN09} who explored this M$_*-$M$_{\rm halo}$ relationship for BCGs at $0.1<z<0.3$. The model of \citet{MOSTER13} suggest an evolution of the M$_*-$M$_{\rm halo}$ relationship with redshift, while observations do show such a trend yet \citep{BROUGH08}. We discuss this in detail in Section \ref{sec:SAMs}.
%============================================================================================

%***********************************************************************************
\begin{table*}
\begin{minipage}{130mm}
\caption{Median values of the whole sample per redshift bin.}
\label{tab:data}
\begin{tabular}{@{}lcccc}
\hline
Redshift bin$^a$ & log M$_{\rm halo}$ [M$_{\odot}$] & log M$_{\rm halo}$ [M$_{\odot}$] 
&log M$_*$ [M$_{\odot}$] &log M$_*$ [M$_{\odot}$] \\
&at observed $z$ &at $z=0^{b}$&at observed $z$ &at $z=0^{b}$  \\        
\hline
High-$z$ ($0.17 < z \leq 0.27$)&13.83&13.93&11.293&11.286\\
Low-$z$   ($0.09 \leq z \leq 0.17$)&13.57&13.63&11.174&11.169\\
\hline
\medskip
\end{tabular}\\
\textbf{$^a$} Both redshift samples (High-$z$  and Low-$z$) are the same size: 441 BCGs.\\
\textbf{$^b$} BCG M$_*$ are corrected by passive evolution, and M$_{\rm halo}$ are corrected to the mass they are likely to have at $z=0$.
\end{minipage}
\end{table*}
 %***********************************************************************************
 \begin{table*}
 \begin{minipage}{130mm}
 \caption{Median BGG/BCG M$_*$ per redshift sample corresponding to the matched M$_{\rm halo}$ groups and clusters. } 
 \label{tab:match}
 \begin{tabular}{@{}lccc}
  \hline
 Redshift   bin& Median $z$& log M$_*$ [M$_{\odot}$]&log M$_*$ [M$_{\odot}$]\\ 
 \hline
 High-$z$ ($0.17 < z \leq 0.27$)&0.136&11.183&11.343\\
 Low-$z$   ($0.09 \leq z \leq 0.17$)&0.214&11.142&11.291\\  
  \hline
  \smallskip
 \end{tabular}
\newpage
\end{minipage}
\end{table*}
%***********************************************************************************
\begin{figure}
\includegraphics[width=\linewidth]{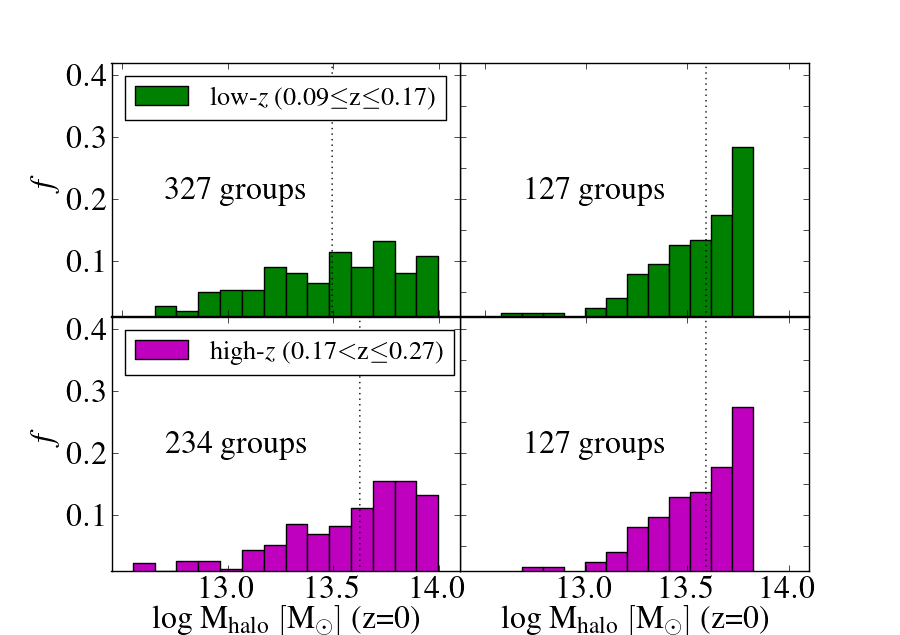}
\caption{Histograms of group halo mass evolved to the present epoch for low (upper panel; $0.09\leq z \leq 0.17$) and high redshift (lower panel; $0.17<z \leq 0.27$) samples. \textbf{Left-hand panels:} M$_{\rm halo}$ distribution (mass evolved to present epochs) for the whole sample per redshift bin. The differences in the median halo masses are illustrated by the dotted lines. \textbf{Right-hand panels:} Matched histograms after selecting the overlap in both redshift samples. Each subsample contains 127 groups, i.e. 127 BGGs. After selecting the groups, from the high-$z$ and low-$z$ samples, with similar halo masses. We guarantee a halo mass like for like comparison between the two different redshift bins.}
\label{fig:g_match}
\end{figure}
%***********************************************************************************
\begin{figure}
\includegraphics[width=\linewidth]{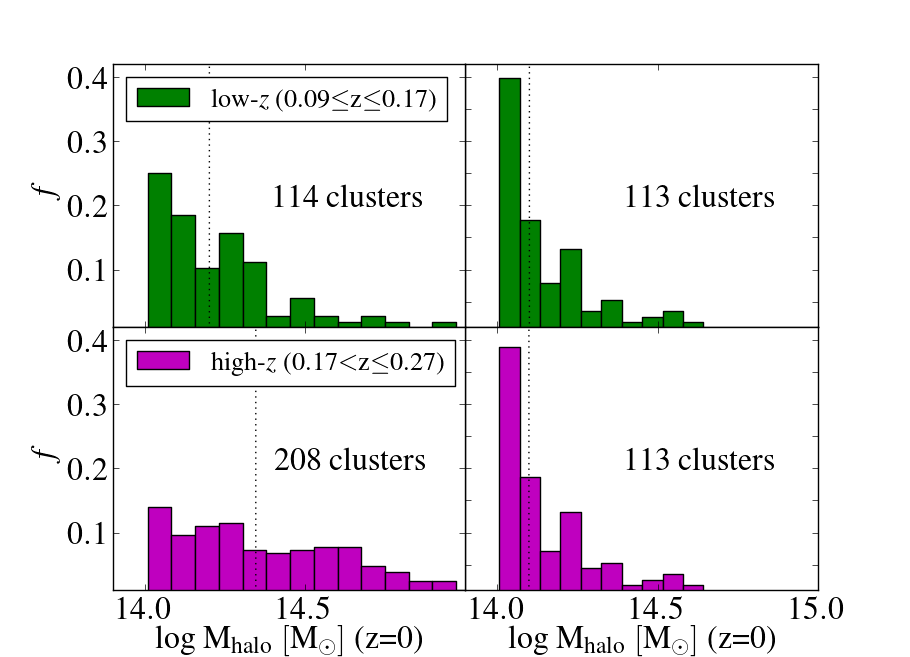}
\caption{Histograms of cluster halo mass evolved to the present epoch for low (upper panel; $0.09\leq z \leq 0.17$) and high redshift (lower panel; $0.17<z \leq 0.27$) samples. \textbf{Left-hand panels:} M$_{\rm halo}$ distribution (mass evolved to present epochs) for the whole sample per redshift bin. The differences in the median halo masses are illustrated by the dotted lines. \textbf{Right-hand panels:} Matched histograms after selecting the overlap in both redshift samples. Each subsample contains 113 clusters, i.e. 113 BCGs. After selecting the clusters, from the high-$z$ and low-$z$ samples, with similar halo masses. We guarantee a halo mass like for like comparison between the two different redshift bins.} 
\label{fig:c_match}
\end{figure}
%***********************************************************************************
\begin{figure}
\includegraphics[width=\linewidth]{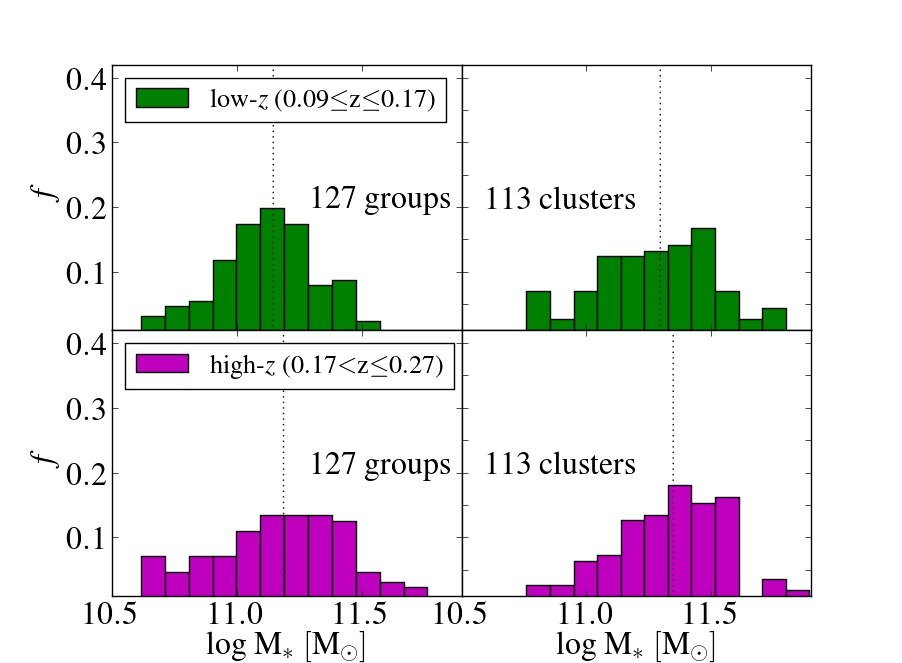}
\caption{Histograms of BGG and BCG stellar mass for low (upper panel; 0.09~$\leq$~z~$\leq$~0.17) and high redshift (lower panel; 0.17~$<$~z~$\leq$~0.27) samples. The median stellar masses are illustrated by the dotted lines. \textbf{Left-hand panels:} M$_{*}$ distribution (corrected for passive evolution to $z=0$) corresponding to the matched group masses. Each redshift bin contains 127 BGGs.  \textbf{Right-hand panels:} M$_{*}$ distribution (corrected for passive evolution to $z=0$) corresponding to the matched cluster masses. Each redshift bin contains 113 BCGs. The median stellar masses are illustrated by the dotted lines. We do not find significant growth in the BGG/BCG stellar mass in the last 3 billion years}
\label{fig:sm_match}
\end{figure}
%***********************************************************************************

\section{BCG growth in the last 3 billion years}
%***********************************************************************************
\begin{figure}
\begin{minipage}[b]{1\linewidth}
\includegraphics[width=1\linewidth]{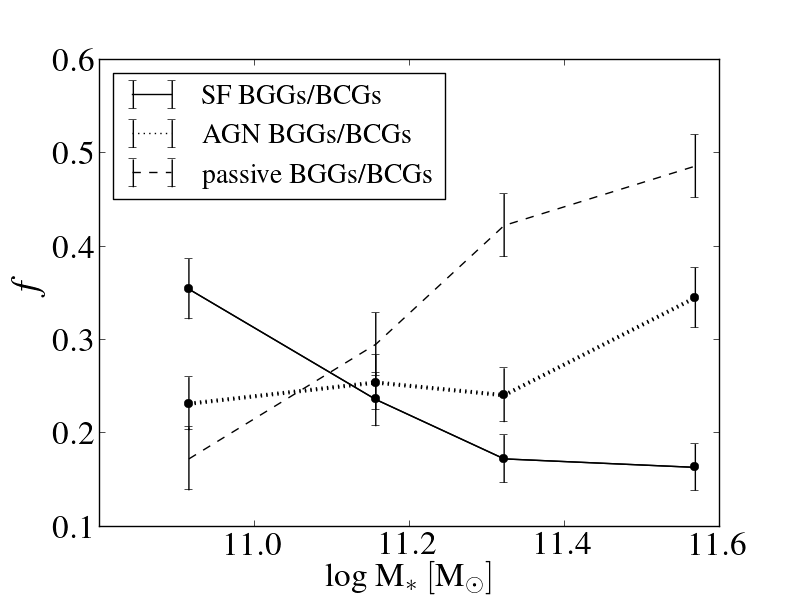}
\end{minipage}
\begin{minipage}[b]{1\linewidth}
\includegraphics[width=1\linewidth]{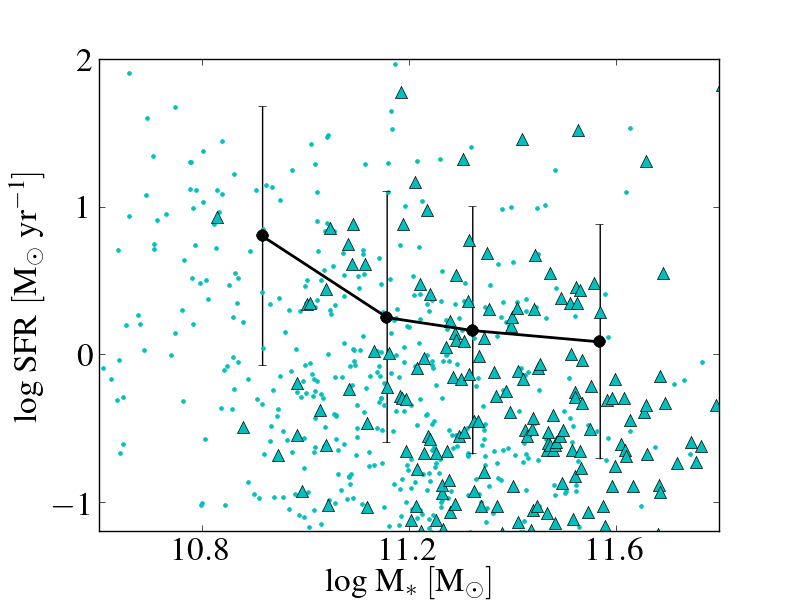}
\end{minipage}
\begin{minipage}[b]{1\linewidth}
\includegraphics[width=1\linewidth]{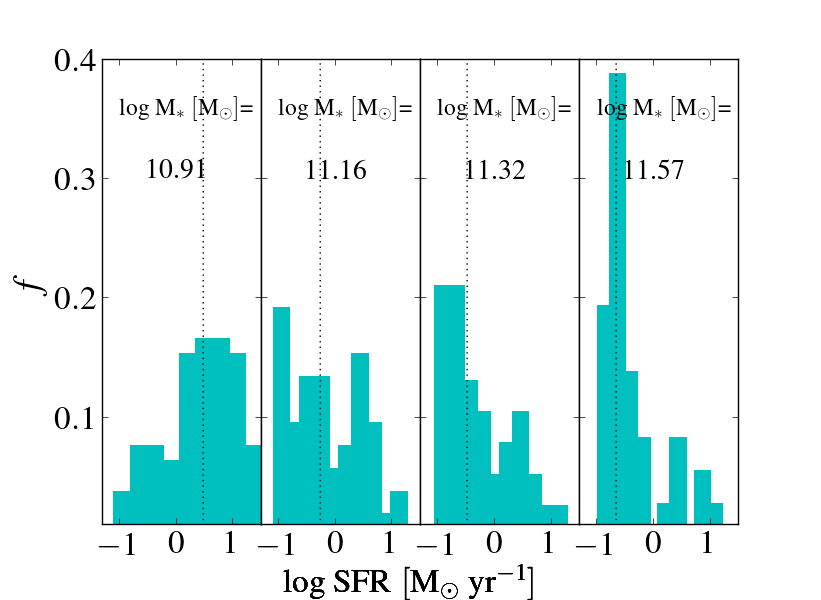}
\end{minipage}
\caption{Examining the activity of BGGs/BCGs as a function of stellar mass. \textbf{Upper panel}: Fraction of ongoing star formation (solid lines), AGN activity (dotted line), and no activity (dashed line) in BGGs/BCGs as a function of M$_{*}$. The fraction of star-forming galaxies decreases with stellar mass, while the fraction of AGNs increases with stellar mass at the high mass end. \textbf{Middle panel}: Median star formation rate of the galaxies that are star forming (235 out of 883 BGGs/BCGs) as a function of stellar mass. The BGGs are shown as circles, and the BCGs as triangles. The SFR decreases with stellar mass. \textbf{Lower panel}: The distribution of the SFR per median M$_{*}$ bin, each histogram corresponds to a point in the middle panel and contains 59 BGGs/BCGs. We show the median SFR as a dotted line. Note how the galaxies gradually quench with increasing stellar mass.}
\label{fig:SFR}
\end{figure}
%***********************************************************************************
%***********************************************************************************
\begin{figure}
\begin{minipage}[b]{1\linewidth}
\includegraphics[width=1\linewidth]{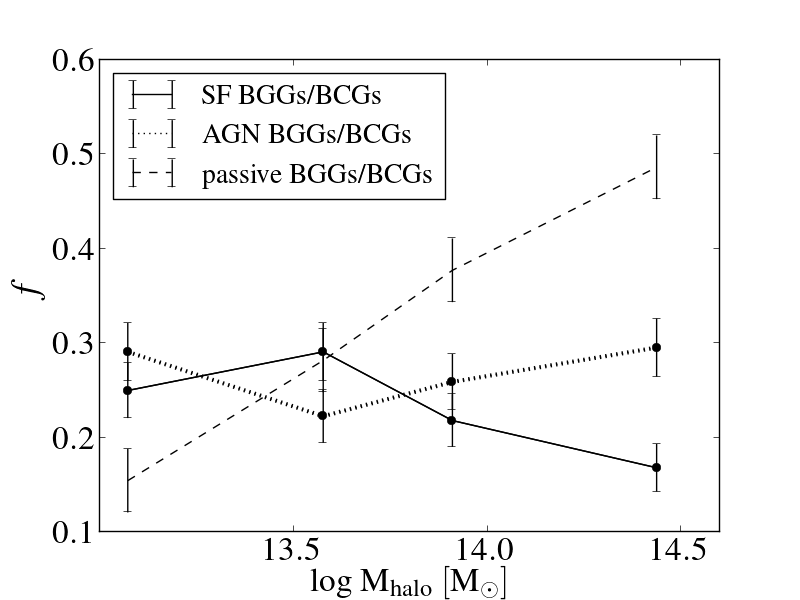}
\end{minipage}
\begin{minipage}[b]{1\linewidth}
\includegraphics[width=1\linewidth]{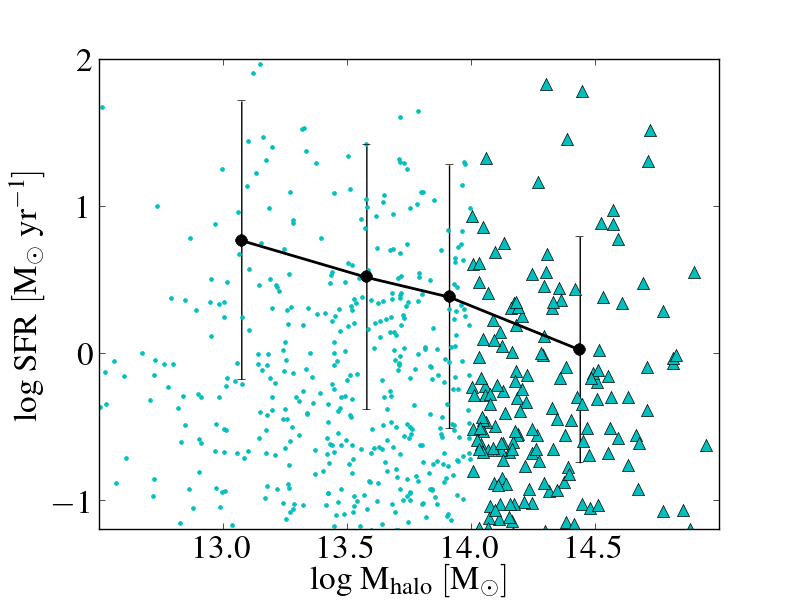}
\end{minipage}
\begin{minipage}[b]{1\linewidth}
\includegraphics[width=1\linewidth, height=6.1cm]{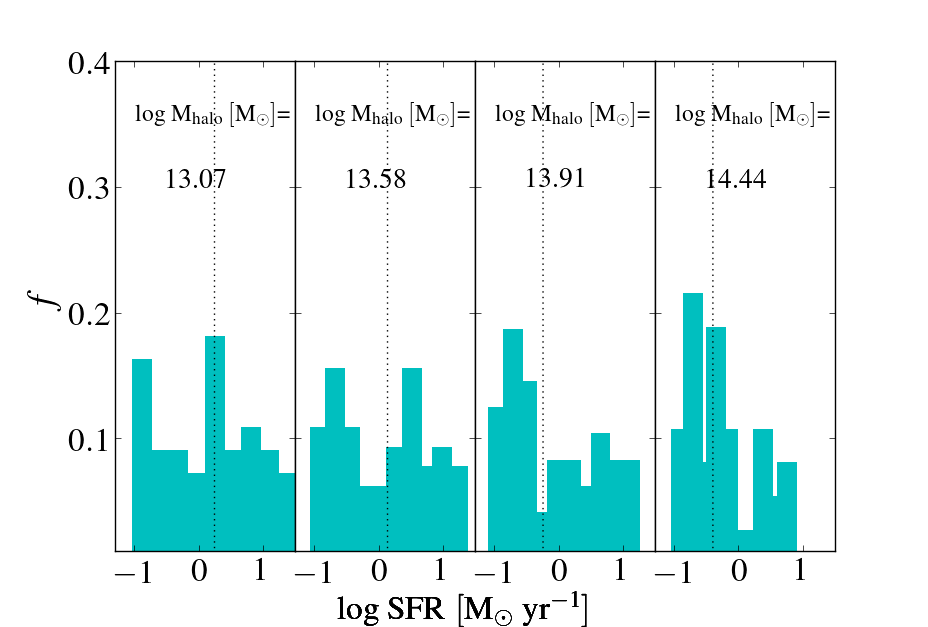}
\end{minipage}
\caption{Examining the activity of BGGs/BCGs as a function of halo mass. \textbf{Upper panel}: Fraction of ongoing star formation (solid lines), AGN activity (dotted line), and no activity (dashed line) in BGGs/BCGs as a function of M$_{\rm halo}$. The fraction of star-forming galaxies decreases with halo mass, while the fraction of AGNs is fairly constant. \textbf{Middle panel}: Median star formation rate of the galaxies that are star forming (235 out of 883 BGGs/BCGs) as a function of M$_{\rm halo}$. The BGGs are shown as circles, and the BCGs as triangles. The SFR decreases with halo mass, although this is likely a result of the stellar mass-halo mass relationship. \textbf{Lower panel}: The distribution of the SFR per median M$_{\rm halo}$ bin, each histogram corresponds to a point in the middle panel and contains 59 BGGs/BCGs. We show the median SFR as a dotted line. There are no clear trends in the distribution of SFR with halo mass.}
\label{fig:SFR_hm}
\end{figure}
%***********************************************************************************
In this paper we are particularly interested in measuring the growth of BCGs in the last 3 billion years. We achieve this by comparing our high-redshift sample with our low-redshift sample. In order to have a reliable comparison between galaxies at higher redshifts and their likely descendants, we need to be sure that we take the growth in halos into account. Here we use the approach implemented in \citet[][]{LIDMAN12}, as described below.  
\\\\
Our halos have been observed at different redshifts so we cannot compare them directly with each other, e.g. halo A at $z=0.25$ is not comparable with halo B observed at $z=0.1$. The same halo A observed later at $z=0.1$ would be very different (larger and more massive). Therefore,  the first step is to evolve all the halos in time to a common redshift, $z=0$, to find the mass that the cluster will likely have by the present day. For this we use the median accretion rate from the model of \citet*{FAKHOURI10}. This model is consistent with other hierarchical structure formation models \citep{WECHSLER02}. 
\\\\ 
As well as the halos, BCGs differ depending on the redshift at which they are observed. We are interested in measuring their growth in stellar mass due to mergers and starburst phenomena, but galaxies also lose material with time due to stellar winds and supernova explosions (mass loss due to the evolution of stars). Here \citep[as shown in][]{LIDMAN12}, we account for this mass loss using the stellar population model of \citet{BRUZUAL03}. Since our redshift range only extends to $z=0.27$, the effect of the stellar mass loss is minimal ($\sim 0.05$dex). In Table \ref{tab:data} we show the median values of the evolved M$_{\rm halo}$ and M$_*$, as well as the M$_{\rm halo}$ and M$_*$ values at the observed redshift. 
\\\\
After having evolved the mass of each halo to the same redshift we can select a set range of halo masses and compare the stellar masses of the BGGs and BCGs from the high and low-redshift samples, having already corrected for the mass loss by passive evolution. For this we examine the M$_{\rm halo}$ distribution of each redshift sample. 
 In the left-hand panels of Figure \ref{fig:g_match} and \ref{fig:c_match}  we show the distributions of the evolved M$_{\rm halo}$ values in each redshift sample for groups and clusters respectively. Note that the distributions vary from one to another: different sample sizes, skewness, and medians. In order to compare the stellar masses of mass-like halos we select all the groups and clusters from the high-$z$ and low-$z$ samples with similar halo masses. The new subsamples are shown in the right-hand panels of Figure \ref{fig:g_match} and \ref{fig:c_match}. They now have the same distributions. This implies  
that the halos in the high redshift subsample are likely to be the ``progenitors" of the halos in the low redshift subsample. 
\\\\
To estimate the M$_*$ growth over the last 3 billion years we calculate the median BGG and BCG M$_*$ corresponding to the M$_{\rm halo}$ matched subsamples. The final M$_*$ distributions are shown in Figure \ref{fig:sm_match}. The left-hand panels are the groups. The right-hand panels are the clusters. The medians are shown in Table \ref{tab:match}. We compute the ratio of the low-$z$~median~M$_*$ to the high-$z$~median~M$_*$. The errors are calculated by bootstrap-resampling of the subsamples. We find no statistically significant growth in the last 3 billion years. The M$_*$ ratio for BGGs and BCGs is $\sim$ 1 within the error bars ($0.92\pm0.07$ for the groups, $0.93\pm0.09$ for the clusters). This result is in agreement with \citet{LIN13}, who found that the BCGs acquire less than $10$ per cent stellar mass between  $0<z<0.5$. We compare this result to the prediction from models in Section \ref{sec:SAMs}. 
%============================================================================================
\section{AGN and Star formation activity}
%***********************************************************************************
\begin{figure}
\includegraphics[width=1\linewidth]{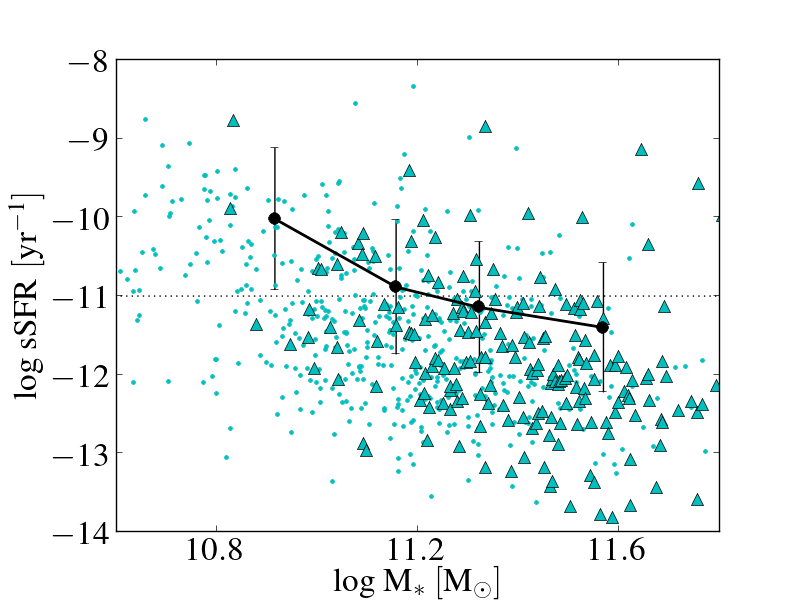}
\caption{Median specific star formation rate of the galaxies that are star forming (235 out of 883 BGGs/BCGs) as a function of stellar mass (M$_{*}$). The BGGs are shown as circles, and the BCGs as triangles. The black points indicate the medians of 59 BGGs/BCGs.The dotted line (sSFR$=-11$ dex) represents the division between active and passive galaxies, defined by \citet{MCGEE11}. A fraction of $0.19\pm0.01$ of the galaxies are above the threshold.} 
\label{fig:sSFR}
\end{figure}
%***********************************************************************************
H$\alpha$ in emission is indicative of star formation and/or AGN activity. To discriminate between the two ionisation sources, we use the \citet*{BPT} diagram described by \citet[][see Section \ref{sec:bpt}]{KEWLEY01}. In previous studies, it has been shown that a large fraction (about 30 per cent overall) of BCGs have signatures of AGN activity, including radio emission \citep{LINDEN07, LIN07,STOTT12}. BCGs are most likely to be in high density environments and comprise old stellar populations. Star formation is not predominate in these galaxies, although it does rarely occur \citep{KAUFFMANN03, KAUFFMANN04, EDWARDS07,ODEA08,BILDFELL08,ODEA10,PIPINO09,LIU12,THOM12}. \citet{LIU12} explored how efficient this star formation is in the stellar mass accretion of BCGs, with a large sample of H$\alpha$ line-emiting BCGs from the SDSS, and found that the star formation is not large enough to contribute to more than 1 per cent of the total stellar mass.  
\\\\
We use the GAMA spectra to identify the activity in our BGGs/BCGs sample and quantify the star formation rates. 236 BGGs/BCGs ($27\pm1$ per cent of the whole sample) have AGN, and 235 ($27\pm1$ per cent of the whole sample) are forming stars. It is not possible to determine the activity level for 412 ($46$ per cent of the whole sample) galaxies as they do not have the necessary emission lines. Of these, 107 show H$\alpha$ line emission ($12$ per cent of the whole sample). These galaxies are likely to be either AGN or star forming, however without the necessary lines it is not possible to distinguish between these possibilities. The remaining 305 galaxies (34 per cent of the whole sample) are passive. In the remainder of this section we focus on the BGGs/BCGs that have been confirmed to either have AGN, star formation or are passive (i.e. we exclude this systems for which we are uncertain).  We distinguish between BGGs and BCGs based on the halo mass of their host systems.  BGGs generally have stellar masses of $10^{10.6}<\,$M$_*<10^{11.6}\,$M$_{\odot}$ while BCGs are generally significantly more massive ($10^{11.2}<\,$M$_*<10^{11.8}\,$M$_{\odot}$).
\\\\  
In the upper panel of Figure  \ref{fig:SFR}  we show the fractions of BGGs/BCGs that have star formation, AGN or are passive as a function of their stellar mass. Both BGGs and BCGs follow a trend of decreasing fraction of star-forming galaxies with increasing stellar mass as seen in other galaxy samples \citep[e.g.][]{KAUFFMANN03,WIJESINGHE11}. In contrast, the fraction of AGN remains fairly constant ($\sim0.25$) for the BGGs, and increases with stellar mass for the BCGs. 
\\\\
We examine the star formation rates (SFRs) of those BGGs/BCGs that show star formation (235 BGGs/BCGs). Their star formation rates extend from our detection limit of $0.1\,$M$_{\odot}$yr$^{-1}$ up to the maximum we reliably measure of $100\,$M$_{\odot}$yr$^{-1}$. However the median SFRs as a function of stellar mass are of the order of $1\,$M$_{\odot}$yr$^{-1}$. In the middle panel of Figure \ref{fig:SFR} we show the median SFR as a function of median M$_*$. In this Figure it is clear that BGGs are the galaxies that are actively star-forming. The BCGs show little to no star formation. In the bottom panel of Figure \ref{fig:SFR} we show the distribution of SFR in each of the M$_*$ bins; each stellar mass bin contains 59 BGGs/BCGs. Note the strong inverse relationship between median SFR and stellar mass. The least massive bin (median M$_*=10^{10.91}\,$M$_{\odot}$) is dominated by the BGGs and presents the galaxies with the highest median SFR. The most massive bin (median M$_*=10^{11.51}\,$M$_{\odot}$) predominately comprises BCGs with very low SFRs ($<1\,$M$_{\odot}$yr$^{-1}$). 
 \\\\
 Figure~\ref{fig:SFR_hm} is equivalent to Figure~\ref{fig:SFR}, but as a function of halo mass. The upper panel shows that the fraction of AGN is constant with halo mass, as it is with stellar mass. In contrast, the fraction of star-forming BGGs/BCGs does not vary as strongly with halo mass, as it does with stellar mass, suggesting that star formation in BCGs is more likely to be dependent on stellar mass rather than environment, like the broader GAMA population \citep{WIJESINGHE12}. The fraction of passive galaxies increases with halo mass similarly to stellar mass. The middle panel shows the SFR as a function of halo mass. The SFR decreases as a function of increasing halo mass; we examine this further below. The bottom panel, shows the distribution of the SFR for each bin in the middle panel. The distribution of SFR does not change with halo mass as it does for stellar mass, further suggesting that in groups and clusters, stellar mass (rather than environment) seems to be driving the SFR relationships.
 \\\\ 
We can examine further the SFRs of the BGGs/BCGs using the specific SFR (sSFR). This is a strong indicator of the star formation evolution of the galaxy, correlating current with previous SFR. In Figure~\ref{fig:sSFR} we show the sSFR as a function of stellar mass for the BGGs/BCGs that are star-forming in our sample. A value of sSFR$=1\times10^{-11}\,$yr$^{-1}$ implies that the galaxy would take 10 Hubble times ($10^{11}$ years) to produce as much mass as it currently has. This means that galaxies with lower sSFRs formed more than 90 per cent of its mass in a single burst that ended when the universe was less than 10 per cent of its current age (i.e. at z>4.5). These galaxies are therefore currently passive compared to their previous star formation. Higher sSFRs imply that galaxies are currently more active \citep[e.g.][]{MCGEE11}. We find that $19\pm1$ per cent of the star-forming BGGs/BCGs are active, and $81\pm1$ per cent are passive. The relationship between sSFR and halo mass is not shown as it is similar to, but weaker than the relationship between sSFR and stellar mass and is likely driven by the stellar mass-halo mass relationship.
\\\\ 
In summary, BGGs are not completely inactive, while most of the BCGs have been shown to be passive galaxies. The percentage of activity, either AGN or star formation, out of our large GAMA sample is 54 per cent, but this is a lower limit owing to the 12 per cent of galaxies which have H$\alpha$ emission but for which we cannot distinguish between AGN and star formation. Nevertheless, the average star formation rates are low ($<10\,$M$_{\odot}\,$yr$^{-1}$). This is consistent with the results of \citet{LIU12} who found that the average star formation in 120 BCGs from SDSS (0.1 $<$ z $<$ 0.4) contributes to less than 1 per cent of their stellar mass. 
%============================================================================================

\section{BGG/BCG position within its host halo}
\begin{figure}
\includegraphics[width=1\linewidth]{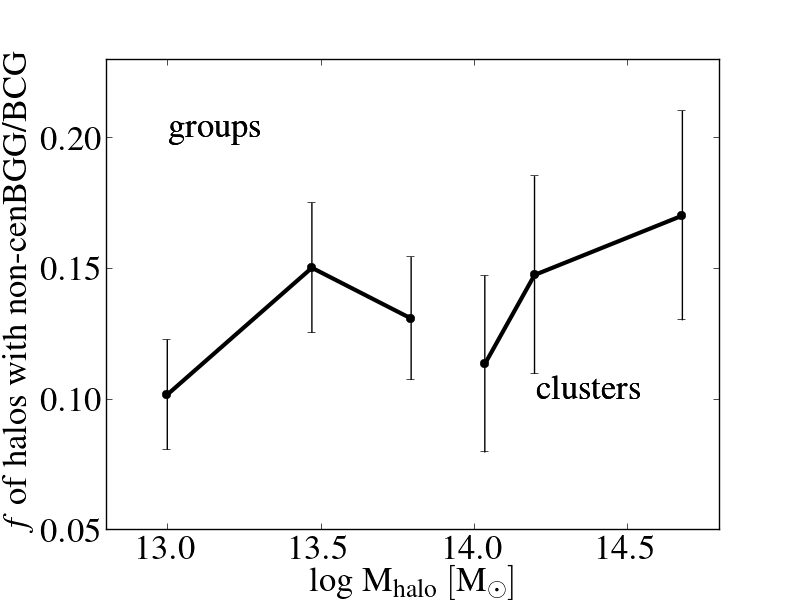}
\caption{Fraction of halos where the BGG/BCG is not centrally located, shown as a function of halo mass. The fraction of non-centrally located BGGs/BCGs increases with increasing halo mass.} 
\label{fig:frac}
\end{figure}

The position of the brightest galaxy in a group or cluster does not always correspond to the centre of the potential well \citep{BEERS83, ZABLUDOFF93, LAZZATI98, LIN04, LINDEN07,PIMBBLET06,COZIOL09,SKIBBA11}. In our total sample we find that $13\pm 1$ per cent of the BGGs/BCGs do \textit{not} lie at the centre of the dark matter halo potential well (hereafter non-central BGGs/BCGs), i.e. 117 BGGs/BCGs out of 883. This is consistent with predictions made by SAMs \citep[$f\sim 0.15$,] []{CROTON06, LOFARO09} and slightly smaller than that measured by \citet[][$f\sim 0.25$]{SKIBBA11} in a large sample of $\sim 2200$ groups ($N\geq 4$). We will further discuss such results in section \ref{sec:cen}.
\\\\
The non-central BGGs/BCGs can be found anywhere between 80 kpc and $\sim1$Mpc away from the halo centre. We note that the two subsamples, central and non-central BGGs/BCGs  will be contaminated by objects in the other subsample, since the iterative centre of light method of determining the halo centre is correct only 90 per cent of the time (see section \ref{sec:group}).  We have used the GAMA mock catalogues to test the level of contamination. While the mock catalogues overestimate the fraction of BCGs offset from the true halo centre compared to the observations, they do provide an estimate of the fraction of contamination of the non-central BCG sample, which is $\sim 1/3$. The impact of the cross-contamination is to dilute the differences that are observed. The real differences between central and non-central BCGs are therefore likely to be stronger than we report.
\\\\
The fraction of non-central BGGs/BCGs varies with M$_{\rm halo}$ (Figure \ref{fig:frac}). In agreement with previous studies, more massive halos are more likely to host a non-central BCG \citep{COZIOL09,SKIBBA11}. Since we do not find significant differences between groups and clusters, in this section we discuss the differences between central and non-central for the whole sample of BGGs/BCGs.

%***********************************************************************************
%***********************************************************************************

%============================================================================================

\subsection{How different are central from non-central BGGs/BCGs? }
\begin{figure*}
\includegraphics[width=12cm]{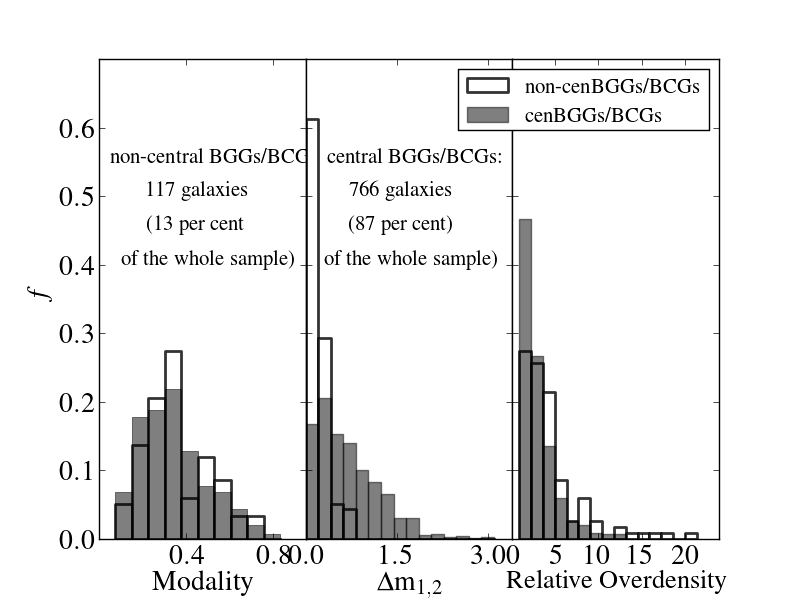}
\caption {Normalised distributions of modality (left-hand panel), dominance (central panel), and relative overdensity (right-hand panel) for central and non-central BGGs/BCGs (shaded and open bars respectively). The non-central BGGs/BCGs are in of halos with lower dominance values and higher relative overdensities. Low dominance and high relative overdensities both suggest the possibility of recent halo-halo mergers.}
\label{fig:hist}
\end{figure*}

\begin{figure}
\includegraphics[width=\linewidth]{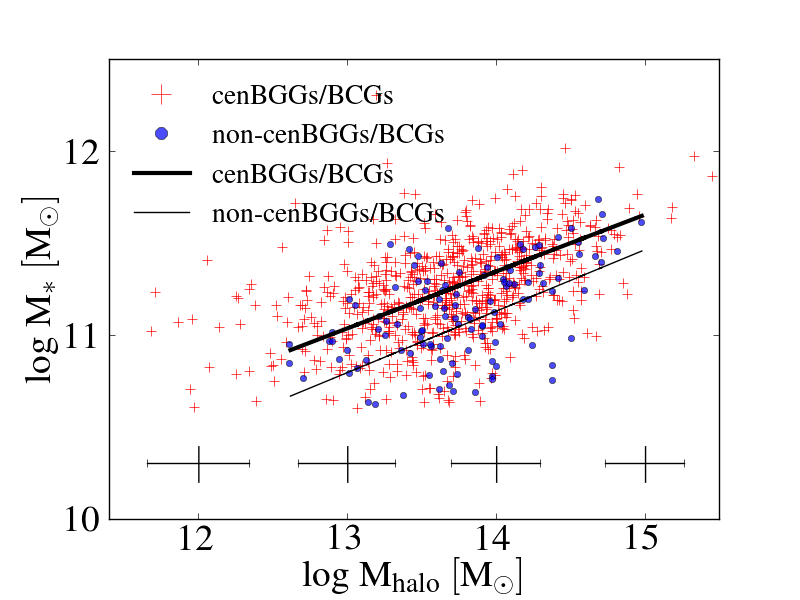}
\caption{BGG/BCG stellar mass - host halo mass relationship for central BGGs/BCGs (red crosses) and non-central BGGs/BCGs (blue circles). The thick line represents the best fit for central BGGs/BCGs. The thin line represents the best fit for non-central BGGs/BCGs. The central and non-central BGGs/BCGs stellar mass - host halo mass relationship are offset in stellar mass by 0.3~dex (a factor of~2) for a given halo mass. }
\label{fig:mass}
\end{figure}

\begin{figure*}
\includegraphics[width=11cm]{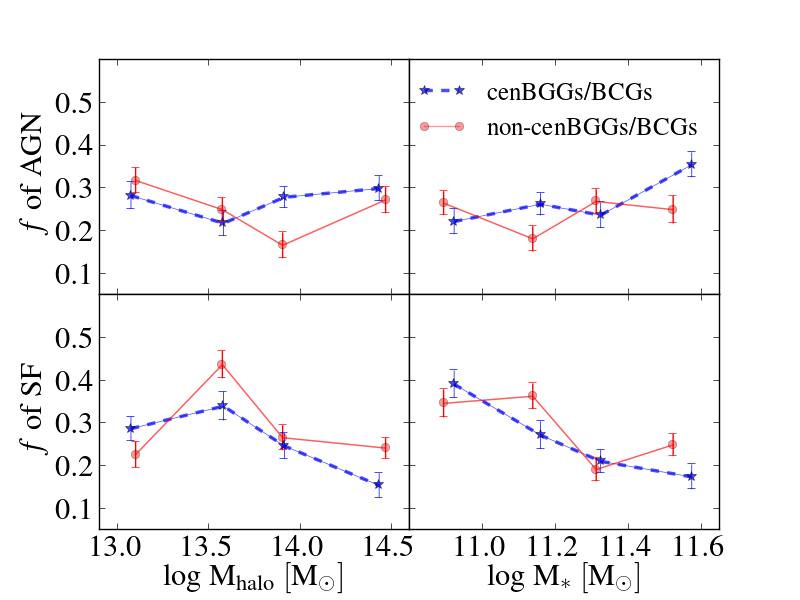}
\caption{\textbf{Left panel}: Fraction of AGNs (upper panels) and star-forming galaxies (lower panels) for central and non-central BGGs/BCGs (red solid line and blue dashed lines respectively) as a function of M$_{\rm halo}$ (left-hand panels) and M$_*$ (right-hand panels). The points represent bins of equal galaxy numbers in a specific mass range. These fractions do not show a dependance on BCG position in the halo. }
\label{fig:agn}
\end{figure*}

In order to avoid biases in sample selection, we first need to be sure that the velocity dispersion distributions of the non-central and central BGGs/BCGs halos are similar enough for these two subsamples to be compared. Therefore, we check the properties of the halos using the ``modality''  parameter described in Section \ref{sec:group}. The modality gives information about the Gaussianity of the velocity dispersion distribution in the halo. Halos with modality $\sim 0.33$ have Gaussian velocity distributions and can be considered to be relaxed. In the left-hand panel of Figure \ref{fig:hist} we compare the modality distributions between the central (shaded) and non-central BGGs/BCGs (open) with a Kolmogorov-Smirnov test (hereafter K-S test). We find a probability of 52 per cent for the two sub-samples to be similar in their Gaussianity. This means that the central and non-central BGGs/BCGs are drawn from similar parent groups. We can now analyse the properties of the galaxies and their host halos in these different subsamples.
\\\\  
The amplitude of the luminosity gap between the BGG/BCG and the second brightest galaxy in the halo (here refer to as the dominance, see section \ref{sec:group}), is expected to be a function of both the formation epoch and the recent infall history of the halo. Small magnitude gap ($\Delta$m$_{1,2}<1$) indicates a recent halo merger, and larger gaps ($\Delta$m$_{1,2}>1$), common in \textit{fossil} groups, is perhaps indicative of a cluster or groups that has not undergone a recent merger. BCGs are expected to be located in clusters with large luminosity gaps \citep[e.g.][]{TREMAINE77, LOH06, SMITH10}. In our entire sample we observe a broad range of dominance ($0< \Delta$m$_{1,2}<3.1$; Figure \ref{fig:hist}), having a long tail towards the higher values. We find a fraction of $20\pm 11$ per cent to be $\Delta$m$_{1,2} >1$ and $3\pm 4$ per cent to be $\Delta$m$_{1,2} >2$. \citet{SMITH10} analysed the dominance of a sample of 59 massive galaxy clusters ($10^{14}$ to $10^{15}\,$M$_{\odot}$). They also found that the distribution of $\Delta$m$_{1,2}$ peaks close to zero and then decays with $\Delta$m$_{1,2}$. They found a fraction of $0.37\pm 0.08$ of their sample had $\Delta$m$_{1,2} >1$ and $0.07\pm 0.05>2$. Our results are consistent with their findings despite the lower average mass of our sample. 
 \\\\ 
 If central and non-central BGGs/BCGs are going through different processes of evolution, this should be reflected in the $\Delta$m$_{1,2}$ values. In the central panel of Figure \ref{fig:hist} we show the $\Delta$m$_{1,2}$ distributions for central BGGs/BCGs (shaded), and non-central BGGs/BCGs (open). From a K-S test we find the central and non-central BGGs/BCGs to have significantly different distributions (probability $<0.01$ of being drawn from the same parent population). Non-central BGGs/BCGs have smaller $\Delta$m$_{1,2}$ which suggests that they reside in halos which are more likely to have undergone a recent halo merger. This is consistent with a naive merger model in which the new system contains two massive bright galaxies, that with time will merge into one. 
  \\\\ 
 To further test this hypothesis, we analyse the relative halo overdensity, which refers to the number of objects surrounding the halo within a given comoving cylinder (see Section \ref{sec:group}), as a proxy for the isolation of the halo. A K-S test performed on these data gives a probability of $<0.01$ for the two subsamples to be similar, which implies that the central and non-central BGGs/BCGs come from different overdensity distributions. The halos with non-central BGGs/BCGs are, on average, part of larger systems (Figure \ref{fig:hist} right-hand panel). This would increase the chances of groups or galaxies falling into the group or cluster.
 \\\\
 In Figure \ref{fig:mass} we show the M$_*-\,$M$_{\rm halo}$ relationship for central (red crosses) and non-central (blue circles) BGGs/BCGs.
 We find that both subsamples follow the same power law within the error bars ($\sim 0.32\pm 0.2$). Both subsamples grow at the same rate as a function of M$_{\rm halo}$. However, they are offset in stellar mass. The central BGGs/BCGs are generally more massive than the non-central BGGs/BCGs ($\sim 0.3$ dex, i.e. on average two times more massive) for a given halo mass. This is also consistent with the naive merger model where the new halo contains the combined mass of both halos, but the merger of the two dominant galaxies is yet to take place. 
 \\\\
 \citet{STOTT12} analysed the BCG luminosity as a function of the BCG offset from the centre of the cluster, finding little correlation between these two properties (power-law index of $0.09\pm0.05$). However, in our analysis we have taken a different approach to  \citet{STOTT12}. We have fixed the halo mass and compared the central and non-central BGGs/BCGs without taking into account the degree of spatial offset. This suggests that difference in properties is a sharp function of whether the BCG is at the centre of light or not.  
\\\\
We also analyse whether the AGN and SF activity of the central BGGs/BCGs is different to that of the non-central BCGs. The feedback prescriptions implemented in SAMs assumes that AGN are hosted in central galaxies only. Therefore, we would expect more AGN activity in central BGGs/BCGs. However, We find that the fraction of AGN activity and star forming galaxies does not differ between the central and non-central BGGs/BCGs implying that neither form of activity is environment-depend for the galaxies in our sample. We illustrate this in Figure \ref{fig:agn}, where the fraction of AGN and star-forming BGGs/BCGs are shown as a function of M$_{\rm halo}$ and M$_*$, respectively. Both subsamples  (central and non-central BGGs/BCGs) follow similar trends: the fraction of AGN remains constant while that of the SF decreases with stellar mass.  This suggests that neither form of activity depends on environment for the galaxies in our sample.
%***********************************************************************************

%***********************************************************************************
%============================================================================================

\section{Discussion}

%***********************************************************************************
%***********************************************************************************

%***********************************************************************************

\begin{figure*}
\includegraphics[width=11cm]{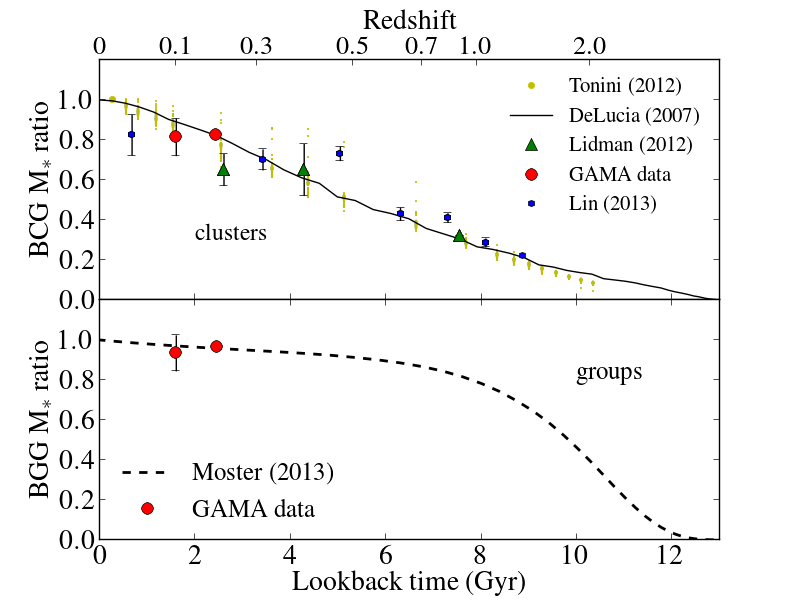}
\caption{Stellar mass ratio evolution with cosmic time. Comparison between observations and hierarchical structure formation models. \textbf{Upper panel}: Comparing the observations presented here (red circles) with the semi-analytic models of \citet[][black line]{DELUCIA07} and \citet[][yellow dots]{TONINI12}. We also show observations from \citet[][green triangles]{LIDMAN12} and \citet[][blue dots]{LIN13}. Each of the observations are normalised to \citet{DELUCIA07}'s model by fixing the  highest redshift point to the model. Our errors were estimated through bootstrapping. The observations follow similar stellar mass growth as the one predicted by the SAMs at high redshifts ($z>0.3$), however this is not the case at lower redshifts. There is a tendency for the low redshift point in each sample to lie below the model predictions. The observations suggest little to no stellar mass growth at $z<0.3$ in contrast with the continuing growth predicted by the models. \textbf{Lower panel}: Comparing our BGG observations (red circles) to the abundance matching model of \citet[][dashed line]{MOSTER13} for the group mass range (median M$_*=10^{11}\,$M$_{\odot}$, median M$_{\rm halo}=10^{13}\,$M$_{\odot}$). The errors were estimated through bootstrapping. Our results agree with the model predictions within error bars.}
\label{fig:growth}
\end{figure*}
%***********************************************************************************

\subsection{Comparison with galaxy formation and evolution models}\label{sec:SAMs}
We have presented observations of 883 brightest groups and clusters galaxies taken from the GAMA survey \citep{DRIVER11}. The sample contains groups and clusters with multiplicities~$>5$ covering a halo mass range of $10^{12}<\,$M$_{\rm halo}<10^{15}\,$M$_{\odot}$. We find an index in the power law of the M$_*-\,$M$_{\rm halo}$ relationship of $b=0.32\pm 0.09$. This is in agreement with \citet{LIN04} who found L$_{K}\propto\,$M$^{0.26}_{200}$, \citet[][]{BROUGH08} who found  L$_{K}\propto\,$M$^{0.24}_{200}$, and \citet{HANSEN09} who found L$_i\propto\,$M$^{0.3}_{200}$. However, \citet{STOTT12} found a much steeper  M$_*-\,$M$_{\rm halo}$ relationship ($b=0.78\pm0.07$) from an X-ray selected sample of BCGs at $z < 0.3$. This is similar to the power-law found by \citet{LIDMAN12}, i.e.~$0.63\pm 0.07$ over a broader redshift range, $0.05\leq z\leq 1.6$. The discrepancies between the power-law values in each analysis could be redshift dependent, but could also be the result of the different methods used in the estimation of galaxy luminosity/mass and halo mass as well as the variety of fitting methods employed.
\\\\ 
 \citet{MOSTER13} have used an abundance matching model, statistically constrained by observations, to predict the M$_*-\,$M$_{\rm halo}$ relationship since $z\sim 4$. They predict the slope of the relationship to be redshift dependent. The relationship we observe is consistent with their prediction in the high stellar mass range ($10^{10.5}$ to $10^{12}\,$M$_{\odot}$ for $z<0.3$). However, the degree of evolution that they predict with redshift depends on observational uncertainties in the stellar mass values, which highlights the effect of the systematics in these kind of measurements.
\\\\
Analysing the growth in the stellar mass of our BGGs/BCGs, we find no significant growth between $z=0.27$ and the present day, for either groups and clusters. We compared the median stellar mass corresponding to the median redshifts of our subsamples ($\bar{z}=0.136$ and $\bar{z}=0.214$), finding a stellar mass ratio of $0.92\pm0.07$ for the groups, and $0.93\pm0.09$ for the clusters. Our group results are in agreement with those predicted by \citet{MOSTER13}. They proposed mass dependant evolution, depending on the star formation efficiency. For the median BGG stellar mass and group mass range (median M$_*=10^{11}\,$M$_{\odot}$, median M$_{\rm halo}=10^{13}\,$M$_{\odot}$) our results agree remarkably well. Our cluster results are consistent with \citet{LIN13} who analysed the growth of BCGs in clusters from the \textit{Spitzer} IRAC Shallow Cluster Survey (ISCS), with halo masses between $(2.5-4.5)\times 10^{14}\,$M$_{\odot}$. They found slow growth at redshifts $z<0.5$ (less than 10 per cent), with more rapid growth (a factor of 2.3) at high redshifts $0.5<z<1.5$. These observational results are in agreement with the SAMs \citep{DELUCIA07, TONINI12} at higher redshifts, finding some differences at lower redshifts. 
\\\\
SAMs predict that a BCG acquires $\sim 30$ per cent of its stellar mass since $z=0.3$, and more than 10 per cent between the median values of our two redshift bins ($\bar{z}=0.136$ and $\bar{z}=0.214$). Our cluster results overall show no growth in this redshift range. Nevertheless, a 10 per cent growth can not be completely ruled out, given that our error bars would allow a maximum of 9 per cent growth for the BCGs and 7 per cent growth for BGGs.  
\\\\ 
We illustrate the comparison between our results and other authors in Figure \ref{fig:growth}. In this Figure we show the BGG (lower panel) and BCG (upper panel) stellar mass ratio evolution over cosmic time. In the upper panel we compare our cluster observations (red circles) with the cluster observations of \citet[][green triangles]{LIDMAN12}, and \citet[][blue dots]{LIN13}, and the SAMs of \citet[][black line]{DELUCIA07} and \citet[][yellow dots]{TONINI12}. In order to compare observations with models, we normalise the highest redshift point of each observational data set to the \citet{DELUCIA07} model. This is justified by the conclusions of \citet[][]{LIN13} whose BCGs were consistent with the models at $z > 0.5$ but become increasingly inconsistent at lower redshifts. The most appropriate way to normalise each observational sample would be to normalise each observational sample at each redshift.  However, the analysis presented in these papers are different, to make a direct comparison between each observation and the model is the most appropriate way of comparison. The BCGs are observed to acquire their stellar mass rapidly from $z = 1.5$ to $z > 0.3$, in agreement with the model predictions. In contrast, below $z \sim 0.3$, the models predict continuing BCG growth that is not observed. The lowest redshift point in all three samples lies below the model curve. This effect of fast growth at high redshifts is also seen in massive field galaxies \citep[e.g.][]{CONSELICE07}.
\\\\
In the lower panel of Figure \ref{fig:growth} we compare our group observations (red circles) with the predictions from the abundance-matching model of \citet[][dashed line]{MOSTER13}. We normalise the highest redshift point of the observations to the model. This model is consistent with our low-redshifts observations. Unfortunately, groups data are not available at higher redshifts, this does not allow us to draw any conclusions on BGG stellar mass growth. 
\\\\
BCG mass growth is observed to be much slower at low redshifts than models predict. The discrepancy with the models suggest that there is some factor in their growth that is not being accounting for. BCGs have grown mainly through mergers \citep[e.g.][]{LIDMAN13} rather than star formation. While models do take into account the timescales for galaxies to merge, they do not always take into account the efficiency of that merging. The efficiency may also evolve with time such that a higher fraction of merging galaxies break-up to become part of the intracluster light at low redshifts \citep[e.g.][]{CONROY07,PUCHWEIN10}. This would result in less mass being added to BCGs in mergers at low redshifts. However, more observations are required to confirm this hypothesis.
%------------------------------------------------------------------------------------------------------------------------------------------------------------------
\subsection{AGN/SF activity in BGGs and BCGs}\label{sec:SF}
At least 27 per cent of the galaxies in our GAMA sample are found to be actively star-forming and another 27 per cent are found to be AGN \citep[classified optically with the BPT diagram of ][]{KEWLEY01}. The fraction of BGGs that host AGN remains fairly constant ($0.25$) with stellar mass. This fraction increases slightly at the high stellar mass end probed by BCGs, but not significantly. This is consistent with \citet{STOTT12} who studied the same stellar mass range in X-ray selected BCGs and found that the fraction hosting radio-loud AGN is constant with stellar mass. We also observe that the AGN fraction is constant with halo mass, showing no environmental dependence above our lowest halo mass of M$_{\rm halo}\sim10^{13}\,$M$_{\odot}$. This contrasts with Stott et al. (2012) who found an increase of the fraction hosting radio-loud AGN with increasing halo mass (from $f\sim0.1$ at M$_{\rm 500}\sim10^{13.9}\,$M$_{\odot}$ to $f\sim0.38$ at M$_{\rm 500}\sim10^{14.7}\,$M$_{\odot}$).
\\\\
The fraction of star-forming galaxies decreases (from $0.4$ to $0.16$) with increasing stellar mass for both groups and clusters, while is fairly constant with halo mass. We analyse the SFR in the BGGs/BCGs that are star-forming and we find that the median SFR is higher ($\sim 8\,$M$_{\odot}\,$yr$^{-1}$) in the less massive galaxies ($10^{10.6}\,$M$_{\odot}<$M$_*<10^{11}\,$M$_{\odot}$), which are mostly BGGs, than in the more massive galaxies which are mostly BCGs ($\sim 1\,$M$_{\odot}\,$yr$^{-1}$). This is consistent with the studies of other galaxy populations \citep[e.g][]{WIJESINGHE12} and the predictions made by the abundance matching model of \citet{YANG13}. They found from a volume-limited sample of BCGs with M$_*>10^{11}\,h^{-1}\,$M$_{\odot}$, that these galaxies are predominately quenched. Meanwhile, galaxies with M$_*<10^{9.5}\,h^{-1}\,$M$_{\odot}$ are forming stars, the galaxies with stellar masses in between show a bimodal distribution of these two groups. We see this bimodality in the BGGs and BCGs from our GAMA sample with median M$_*=10^{11.16}\,$M$_{\odot}$. In contrast, galaxies with masses more than M$_*>10^{11.3}\,$M$_{\odot}$ are mostly quenched (Figure \ref{fig:SFR}). We found that these trends are driven by stellar mass rather than by the host group/cluster environment. Overall $\sim19$ per cent of the star-forming BGGs/BCGs can be identified as active galaxies \citep[sSFR$\,>1\times10^{-11}\,$yr$^{-1}$; ][]{MCGEE11}. Leaving $\sim 81$ per cent as passive galaxies. The active galaxies are mainly BGGs.
\\\\
BGGs are not completely dormant, while BCGs present significantly less star formation but higher fractions of AGN activity. Nevertheless the star formation rates for both BGGs and BCGs are not high enough to contribute significant amount of stellar mass in these giant galaxies. A SFR of $10\,$M$_{\odot}$ per year can be found in the BGGs, but in general most of the BGGs and BCGs have SFR~$<3\,$M$_{\odot}$yr$^{-1}$. Our results agree with those of \citet{LIU12}. However, this fact cannot be overlooked in theoretical work. \citet{TONINI12} showed that by including star formation in SAMs the predicted photometric colours  are significantly improved in terms of reproducing the observations. More specifically, luminosities in the K-band in their model are in better agreement with the observations of \citet{BROUGH08}, \citet{STOTT08}, \citet{WHILEY08}, \citet{COLLINS09}, and \citet{LIDMAN12} than previous models. 
%------------------------------------------------------------------------------------------------------------------------------------------------------------------
\subsection{Central vs non-Central BGGs/BCGs}\label{sec:cen}
We find that 13 per cent of the BGGs/BCGs in the GAMA sample are not centred in their host halo. This fraction is consistent with that predicted in  SAMs \citep[$0.1<f<0.2$;][]{CROTON06, LOFARO09}. In contrast, \citet{SKIBBA11} in a sample selected from the SDSS group catalogue found a larger fraction (0.25 for M$_{\rm halo} \sim10^{12}-10^{13}\,$M$_{\odot}$ to 0.4  for M$_{\rm halo}=5 \times 10^{13}\,$M$_{\odot}$). Despite the difference between our sample and that of \citet{SKIBBA11}, we agree with their overall conclusions, that the fraction of non-central BCGs increases with increasing M$_{\rm halo}$ (Figure \ref{fig:frac}).
\\\\
 After analysing the properties of the halos of the BGGs/BCGs in our sample using the dominance ($\Delta$m$_{1,2}$) and relative overdensity parameters (Figure \ref{fig:hist}), we find that the non-central BGGs/BCGs halos have significantly smaller $\Delta$m$_{1,2}$ values, and higher relative overdensities. In contrast, central BGGs/BCGs halos are shown to have a broader range of values ($0<\Delta$m$_{1,2}<3.4$).  The dominance and overdensity results both suggest that the non-central BGGs/BCGs are likely to be a result of recent halo-halo mergers. This conclusion is further strengthened by the difference in stellar mass between the central and non-central BGGs/BCGs. The non-central BGGs/BCGs have most likely fallen into their current system as the central galaxy of a lower mass system. Dynamical friction will act upon this BGG/BCG, causing it to fall to the centre of its new system. The fact that the fraction of non-central BGGs/BCGs increases with increasing halo mass suggests that the timescale for the BCG to merge with the central galaxy of the other halo is longer. 
%============================================================================================
\section{Summary and conclusions}
We have analysed a large (883 galaxies) and homogeneous sample of low redshift ($0.09<z<0.27$) brightest group and cluster galaxies from the Galaxy And Mass Assembly (GAMA) survey. We summarise our conclusions below.  
\\\\(a) By comparing the BGG/BCG stellar mass in like-with-like halos we find no significant growth over this period of cosmic time. After comparing our results with previous analyses we conclude that BCGs acquire their stellar mass rapidly at early epochs ($z>0.3$). Below redshift $z\sim0.3$ the stellar masses increase more slowly. This is possibly because the timescales or efficiencies for merging evolve. While observations are more consistent with the models (SAMs and abundance-matching models) at higher redshift, there are still small discrepancies at low redshifts. We stress the importance of taking into account the stellar mass~-~halo mass relationship for such a comparisons. 
\\\\
(b) We find that BGGs/BCGs are not completely dormant; at least 27 per cent of our sample host AGN and another 27 per cent are star forming. Their star formation rate decreases with stellar mass, from $10\,$M$_{\odot}\,$yr$^{-1}$ at M$_*\sim 10^{10.8}\,$M$_{\odot}$ to less than 1 M$_{\odot}\,$yr$^{-1}$ at M$_*\sim 10^{11.6}\,$M$_{\odot}$. Therefore, BGGs are actively star-forming while BCGs are mostly quenched with higher fractions of AGN activity.  At stellar masses 10$^{11}\,$M$_{\odot}<\,$M$_*<10^{11.4}\,$M$_{\odot}$  we find a bimodal population of star-forming and quenched systems. We conclude that despite the presence of star-formation in BCGs the SFRs are not high enough for star formation to contribute significantly to the stellar mass growth of these galaxies. 
\\\\     
(c) We also examine the position of the BGGs/BCGs with respect of their dark matter halo and find that around $\sim13$ per cent of the BGGs/BCGs are not centrally located. The halo properties, dominance and relative overdensity, in non-central BGGs/BCGs halos suggest that these halos have undergone recent mergers. This is further proven by the overall stellar mass difference between central and non-central BGGs/BCGs. We suggest that non-central BGGs/BCGs were the central galaxies in a smaller system that fell into the current system not long ago. The fraction of AGNs and star forming galaxies is roughly the same for central and non-central BGGs/BCGs.  
\section{ACKNOWLEDGEMENTS}
We thank the anonymous referee for their thoughtful comments that have improved this paper. GAMA is a joint European-Australasian project based around a spectroscopic campaign using the Anglo-Australian Telescope. The GAMA input catalogue is based on data taken from the Sloan Digital Sky Survey and the UKIRT Infrared Deep Sky Survey. Complementary imaging of the GAMA regions is being obtained by a number of independent survey programs including GALEX MIS, VST KiDS, VISTA VIKING, WISE, Herschel-ATLAS, GMRT and ASKAP providing UV to radio coverage. GAMA is funded by the STFC (UK), the ARC (Australia), the AAO, and the participating institutions. The GAMA website is http://www.gama-survey.org/.
\\\\
C.L. is the recipient of an Australian Research Council Future Fellowship (program number FT0992259).
\\\\
MLPG acknowledges support from a European Research Council Starting Grant (DEGAS-259586).
\\\\
AEB acknowledges the Australian Research Council (ARC) and Super Science Fellowship funding for supporting this work [FS100100065].

%============================================================================================
%%%%%%%%%%%%%%%%%%%%%%%%%%%%%%%%%%%%

% \bsp % ``This paper has been produced using the ...''
\bigskip
%\newpage
\setlength{\bibhang}{2.0em}
\setlength\labelwidth{0.0em}
\bibliographystyle{mn2e}
\bibliography{POliva_2014.bib}
\label{lastpage}

\end{document}